\documentclass[twocolumn]{aastex63}

\usepackage{xcolor}
\usepackage{amsmath}

\submitjournal{ApJ}

\shorttitle{FRB multiwavelength counterparts}
\shortauthors{Chen, Ravi \& Lu} 
\graphicspath{{./}{figures/}}

\begin{document}

\title{The multiwavelength counterparts of fast radio bursts}
\correspondingauthor{Ge Chen}
\email{gcchen@caltech.edu}

\author{Ge Chen}
\affiliation{Cahill Center for Astronomy and Astrophysics, MC 249-17 California Institute of Technology, Pasadena CA 91125, USA}

\author{Vikram Ravi}
\affiliation{Cahill Center for Astronomy and Astrophysics, MC 249-17 California Institute of Technology, Pasadena CA 91125, USA}

\author{Wenbin Lu} 
\affiliation{Cahill Center for Astronomy and Astrophysics, MC 249-17 California Institute of Technology, Pasadena CA 91125, USA}

\begin{abstract}
The engines that produce extragalactic fast radio bursts (FRBs), and the mechanism by which the emission is generated, remain unknown. Many FRB models predict prompt multi-wavelength counterparts, which can be used to refine our knowledge of these fundamentals of the FRB phenomenon. However, several previous targeted searches for prompt FRB counterparts have yielded no detections, and have additionally not reached sufficient sensitivity with respect to the predictions. In this work, we demonstrate a technique to estimate the ratio, $\eta$, between the energy outputs of FRB counterparts at various wavelengths and the radio-wavelength emission. Our technique combines the fluence distribution of the FRB population with results from several wide-field blind surveys for fast transients from the optical to the TeV bands. We present constraints on $\eta$ that improve upon previous observations even in the case that all unclassified transient events in existing surveys are FRB counterparts.  In some scenarios for the FRB engine and emission mechanism, we find that FRB counterparts should have already been detected, thus demonstrating that our technique can successfully test predictions for $\eta$. However, it is possible that FRB counterparts are lurking amongst catalogs of unclassified transient events. Although our technique is robust to the present uncertainty in the FRB fluence distribution, its ultimate application to accurately estimate or bound $\eta$ will require the careful analysis of all candidate fast-transient events in multi-wavelength survey data sets.
\end{abstract}



\section{Introduction} \label{sec:intro}

Fast radio bursts (FRBs) are short ($\sim$ ms) and luminous ($\sim 10^{42}$\,erg\,s$^{-1}$) radio pulses detected at extragalactic distances. There have been nearly a hundred FRBs reported \citep{2016PASA...33...45P}, and the estimated rate is $\sim 10^3 \rm~sky^{-1}~day^{-1}$  \citep{2018MNRAS.475.1427B}. Five FRB sources have been directly associated with host galaxies, revealing a range of galaxy classes, and source environments \citep{2017Natur.541...58C,2019Sci...365..565B,2019Natur.572..352R,2019Sci...366..231P,2020Natur.577..190M}. Repeat bursts have been observed from 20 FRB sources \citep{2016Natur.531..202S,2019Natur.566..235C,2019ApJ...885L..24C,2019ApJ...887L..30K,2020arXiv200103595F}, indicating that at least some FRBs originate from non-catastrophic events \citep[see also][]{2019NatAs...3..928R}.  
The exact FRB emission mechanism(s) and engine(s) remain elusive. 

The high brightness temperatures ($\sim 10^{35}$\,K) of FRBs require a coherent emission process.  Although several astrophysical coherent emission mechanisms are identified with Galactic sources \citep{2017RvMPP...1....5M}, these mechanisms encounter difficulties with the energy scales of FRBs. Two  classes of mechanisms have been proposed for FRBs \citep[although see, e.g.,][]{2020arXiv200102007L}: synchrotron masers \citep[e.g.,][]{2014MNRAS.442L...9L,2017MNRAS.465L..30G,2017ApJ...842...34W,2017ApJ...843L..26B,2019MNRAS.485.4091M} and coherent curvature radiation \citep[e.g.,][]{2016MNRAS.457..232C,2017MNRAS.468.2726K}. Synchrotron masers require a population inversion in the emitting lepton energy and pitch-angle distribution. This is generally thought to be possible in an ultra-relativistic (bulk Lorentz factor $\Gamma\gg1$) radiative shock driven into a significantly magnetized plasma (magnetization parameter $\sigma \gtrsim 10^{-3}$); the shock is mediated by Larmor-rotating charges, which results in the population inversion within the shock. The curvature-radiation mechanism instead scales ideas for the generation of pulsar radio emission to FRB energy scales, invoking coherently radiating bunches of relativistic ($\Gamma\sim30$) leptons accelerated by magnetic reconnection events. These radiation processes are discussed in the context of several progenitor models.  The majority of these models involve highly magnetized neutron stars / magnetars, because the short durations and high luminosities of FRBs require compact, active engines with large energy budgets and emission-region field strengths of $\gtrsim10^{10}$\,G.  

Several classes of FRB models (emission mechanism and/or engine) predict prompt multi-wavelength counterparts, and specify the ratio between the energy emitted by the counterpart and by the FRB. Synchrotron masers initiated by ultra-relativistic shocks are accompanied by synchrotron emission from fast-cooling shock-heated electrons that sweeps through the $\gamma$-ray and X-ray bands on sub-second timescales, or perhaps through the optical/NIR bands in the case of an electron-positron plasma upstream of the shock \citep{2019MNRAS.485.4091M}. Specific luminosities comparable to or greater than the FRB luminosities are predicted for the higher-energy emission. Additionally, although the curvature-radiation mechanism does not naturally produce multi-wavelength emission, the mechanism may be triggered by events that do radiate across the electromagnetic spectrum \citep[e.g., the cosmic comb model;][]{2017ApJ...836L..32Z}. 

We define the ratio between the energy radiated by an FRB event in a given electromagnetic band and in the radio band as \begin{equation}
\label{eqn: eta}
    \eta(\nu_c)=\frac{F_{\rm Band}}{F_{\rm Radio}} \approx \frac{\nu_{1,\rm Band} \cdot F_{\nu,\rm Band}}{\nu_{1,\rm Radio}\cdot F_{\nu \rm Radio} }.  
\end{equation}
Here, $\nu_c$ is the central frequency of the band of interest, $F_{\rm Radio}$ and $F_{\rm Band}$ are the band-integrated fluences in the band of interest and the radio band, respectively, $F_{\nu, \rm Radio}$ and $F_{\nu, \rm Band}$ are the specific fluences, and $\nu_{1, \rm Band}$ and $\nu_{1, \rm Radio}$ are the lower-frequency bounds of these two bands.  The commonly-made approximation in the second step \citep{1997NCimB.112...11G} assumes that the bands span natural-logarithmic frequency intervals, but is accurate in  general when the specific fluence scales as $F_{\nu}\propto\nu^{-2}$, and the band upper-frequency bound $\nu_2$ is much greater than the lower bound $\nu_1$.  We also define a  fluence ratio as 
\begin{equation}
\label{eqn: eta_nu}
\eta_{\nu} (\nu_c) = \frac{F_{\nu,\rm Band}} {F_{\nu,\rm Radio}}.
\end{equation}
In this work, we consider how constraints on $\eta$ and $\eta_{\nu}$ based on the possible detection, or non-detection, of multiwavelength FRB counterparts can test FRB models.  


Until December 2019, no FRB multi-wavelength counterpart has been confirmed.  Most previous observational constraints on $\eta$ are derived from searches for multi-wavelength transient emissions that are close to the FRBs in both time and location (see Section~\ref{subsec:discussion_obs} for references). It is also possible to search for  multi-wavelength transient events that are close to the FRBs in location only, but at any time.

In this work, we explore a third method: a blind search of the whole sky for unclassified multi-wavelength transient events at any time. Several surveys across the optical and high-energy bands explore the sky with sub-second time resolution. We develop and demonstrate a method to estimate $\eta$ by combining relevant multi-wavelength survey parameters with the fluence distribution of the currently observed FRB population. This method can only directly constrain $\eta$ when the statistics of unclassified short-duration transient events are published.  However, these statistics are rarely published.  We therefore compare constraints derived assuming no multi-wavelength FRB counterparts have been detected with predictions from FRB emission models. We find in several scenarios that existing surveys are likely / unlikely to have already detected FRB counterparts. For some surveys, we also consider the case where a fraction of their unclassified events are indeed FRB counterparts to derive upper bounds on $\eta$. 

We propose our method for three reasons.  First, a blind search for counterparts is meaningful since some models predict cases where the radio emission from an FRB is non-detectable while the counterpart is. For example, \citet{2019MNRAS.485.4091M} show that high-energy counterpart emission may escape a dense medium surrounding the source even when the radio emission is subject to the obscuring effects of scattering and absorption.  Second, it is reasonable to make use of the statistical properties of FRBs.  The estimated FRB rate above $F_{\nu,0}\sim 2 \rm~Jy~ms$ is $R_0 \approx 1.7\times 10^3 \rm~~sky^{-1} day^{-1}$  \citep{2018MNRAS.475.1427B}, sufficiently large to be treated as a common events for relatively large telescopes scanning across most of the sky in a blind survey \citep{2016ApJ...824L...9V}.  Third, we will show that our technique provides stronger constraints than previous multi-wavelength observations in the current situation wherein no counterpart has been detected.  

We explain the method to estimate the band-to-radio energy ratio $\eta$ in Section~\ref{sec:methods}, introduce the surveys under consideration in Section~\ref{sec:inst}, and show our calculations and results in Section~\ref{sec:results}.  In Section~\ref{sec:discussion}, we compare our results with theoretical predictions made by leading FRB emission models.  We also compare our results with previous observational constraints, and discuss observational strategies for future blind searching. We conclude in Section~\ref{sec:conclusion}.

\section{Methods} \label{sec:methods}


We adopt the broken power-law specific-fluence cumulative distribution function (CDF) estimated using the Australian Square Kilometre Array Pathfinder (ASKAP) and Parkes FRB samples  \citep{2019MNRAS.483.1342J}:
\begin{subequations} 
\label{eqn:bpl}
\begin{align}
    R(F_{\nu}) &=\int_F^{\infty} r(F_{\nu}') dF_{\nu}',\\
    &= R_0\left( \frac{F_{\nu}}{F_{\nu,0}} \right)^{\alpha_1} ~~~(F_{\nu,\rm min} < F_{\nu} < F_{\nu,b}), \label{eqn:bpl1}\\
    &=R_b\left( \frac{F_{\nu}}{F_{\nu,b}} \right)^{\alpha_2}~~~(F_{\nu} > F_{\nu,b}). \label{eqn:bpl2}
\end{align}
\end{subequations}
Here, $R(F_{\nu})$ is the FRB rate (in the usual units of sky$^{-1}$\,day$^{-1}$) above a given fluence threshold $F_{\nu}$ in the radio band, $r(F_{\nu})$ is the differential fluence distribution function, $\alpha_1 = -1.18$,  $\alpha_2 = - 2.2$, $F_{\nu,\rm min}$ is the (observationally unconstrained) fluence cut-off in the radio-band, $F_{\nu,0}\sim 2$\,Jy\,ms is the fluence completeness threshold for the Parkes FRB searches \citep{2015MNRAS.447.2852K}, $R_0 \approx 1.7\times 10^3$\,sky$^{-1}$\,day$^{-1}$ is the estimated rate  above $F_{\nu,0}$ \citep{2018MNRAS.475.1427B}, $F_{\nu,b}$ is the fluence break which we choose to be 15\,Jy\,ms, and $R_b \approx 171$\,sky$^{-1}$\,day$^{-1}$ is the rate above $F_b$ calculated from Eq.~(\ref{eqn:bpl1}).  

We assume that in any other emission band the fluence CDF, $\tilde{R}$, has the same functional form as $R$, except with a shift in abscissa (i.e., horizontally), and a re-normalization:
\begin{align}
\label{eq: rate}
    \tilde{R}\left(F_{\nu}\right) =  R\left(\frac{F_{\nu}}{\eta_{\nu}}\right).
\end{align}
We use properties of multi-wavelength surveys to estimate $\eta_{\nu}$ by calculating the degree to which the distribution needs to be shifted in its abscissa to achieve the same detection rate in two different bands. 

We now describe how estimates of $\eta$ and $\eta_{\nu}$ are made. Suppose that a transient survey has a field of view (FOV) of $\Omega$ steradians, effectively lasts for $n$ days, and has not detected any FRB counterpart.  The survey operates at frequencies from $\nu_1$ to $\nu_2$, with a center frequency of $\nu_c$. Our method involves the following steps: 
\begin{enumerate}
\item Convert the instrumental detection sensitivity threshold to an energy flux, $f_0$. \label{itm:step_f0} 
\begin{itemize}
\item If the photon flux threshold $f_{\rm ph}$ is specified, we calculate $f_0$ using the specifications of each instrument and the weighted average photon frequency $\langle \nu  \rangle$ in this band, assuming a typical photon index of $-2$ (i.e. a spectral index of $-1$; see e.g. \citealt{2009ApJ...697.1071A})\footnote{This assumption yields larger estimates for $\eta$ than if a steeper photon index was assumed (see Section~\ref{sec:results}). Some previous studies \citep[e.g.,][]{2016ApJ...827...59T} assume steeper photon indices.}: 

\begin{equation}
\label{eqn:E_eff}
\begin{split}
f_0 &= f_{\rm ph} \cdot h \langle \nu \rangle, \\
 &= f_{\rm ph} \cdot h~ \frac{\int^{\nu_2}_{\nu_1} d\nu~\nu^{-2}~\nu}{\int^{\nu_2}_{\nu_1} d\nu~ \nu^{-2}}, \\
    &= f_{\rm ph} \cdot h~\frac{ln(\nu_2 / \nu_1)}{\nu_1^{-1}-\nu_2^{-1}}.
\end{split}
\end{equation}

\item If we know the limiting magnitude $m$, 
\begin{equation}
\label{eqn: f0}
f_0=10^{-0.4 m} \cdot \nu_1 \cdot f_{\nu\rm ,m0}.
\end{equation}
Here, $f_{\nu \rm ,m0}$ is the specific energy flux of an object with zero magnitude in that band and magnitude system \citep{1998A&A...333..231B, 1994AJ....108.1476F,1996AJ....111.1748F}. 
\end{itemize}

\item \label{itm:step_scale} 
Scale the detection limiting energy flux $f_0$ according to a timescale $\Delta t$.  Specifically, if the survey has an automatic self-trigger algorithm for burst candidates, we use the trigger timescale(s)  $t_{\rm trig}$.  Otherwise, we use the nominal instrumental time resolution $t_{\rm res}$. Since the signal to noise ratio ${\rm SNR} \propto \Delta t^{1/2}$, the detection flux threshold $f_0 \propto \Delta t^{-1/2}$.  (Note that if the burst duration $t_{\rm burst} < \Delta t$, the corresponding intrinsic burst flux needs to be higher than $f_0$.  Otherwise, they are the same.) 

\item \label{itm:step_F0} Compute the fluence thresholds of the burst in the band of interest.  
\begin{itemize}
\item If $t_{\rm burst} < \Delta t$, the band-integrated burst fluence limit is $F_{0, \rm Band} = f_{0} \cdot \Delta t$. 
\item Otherwise, $\Delta t$ is too short for the burst.  One should either choose a different timescale or bin adjacent time samples. 
\end{itemize}

The specific fluence threshold is $F_{_\nu, 0, \rm Band} \approx F_{0, \rm Band} / \nu_1$.  In addition, we require that the number of photons received within the timescale ($\Delta t$) by the telescope's effective collecting area ($A$) is at least one.  When this is not satisfied, we replace the photon flux threshold $f_{\rm ph}$ (step \ref{itm:step_f0}) with $1 \rm~photon~\Delta t^{-1} ~A^{-1}$, and repeat the previous steps.

\item \label{itm:step_rate} Calculate the event-rate upper bound in this survey if no candidate were detected: 
\begin{equation}
\label{eqn:rate}
R(F_{\nu,0,\rm Band}) \leq \frac{3}{n} \frac{4\pi}{\Omega} ~\rm sky^{-1}~d^{-1}.
\end{equation}
A non-detection means a Poisson single-sided upper limit of 3 at the 95\% confidence level \citep{1986ApJ...303..336G}.  Alternatively, if there were $x$ candidate events in this survey, the event rate would be 
\begin{equation}
\label{eqn:rate_x}
R = \frac{x}{n} \frac{4\pi}{\Omega} ~\rm sky^{-1}\,d^{-1}. 
\end{equation}

\item \label{itm:step_F0_radio} Solve for the radio band fluence threshold $F_{\nu,0, \rm Radio}$ that would have produced the same rate $R$ using Eqn.\ref{eqn:bpl}. 

\item \label{itm:step_ratios} Find the fluence ratios:
\begin{equation}
    \eta_{\nu}(\nu_c) \leq F_{\nu, 0, \rm Band} / F_{\nu, 0, \rm Radio},
\end{equation}
and 
\begin{equation}
    \eta(\nu_c) \leq F_{0, \rm Band} / F_{0, \rm Radio} \approx \frac{F_{0,\rm Band} / \nu_{1, \rm Band}}{F_{0, \rm Radio} / \nu_{1, \rm Radio}}.
\end{equation}
Here, $F_{\nu,0,\rm Band}$ and $F_{\nu,0,\rm Radio}$ have been found in steps \ref{itm:step_F0} and \ref{itm:step_F0_radio}, respectively. 

\end{enumerate}

\section{Existing Surveys}\label{sec:inst}

\begin{deluxetable*}{c c c c c c}
\tablecaption{Existing surveys and instruments.
\label{ta:ins}}

 \tablehead{
  \colhead{Instrument} & \colhead{Band} & \colhead{Effective duration} & \colhead{Detection threshold} & \colhead{Timescales} & \colhead{FOV} \\ 
 & & \colhead{(days)} & & & 
 }
 
 \startdata
  MAGIC \tablenotemark{a} & 50~GeV--50~TeV & $\approx 1227$ & $7.6\times10^{-12}~\rm photons~cm^{-2}~s^{-1}$ \tablenotemark{k} & e.g. 10 ms & 4.8 sq. deg. \\ 
  \textit{Fermi}/LAT\tablenotemark{b} & 20~MeV--300~GeV & 4132 & $3\times 10^{-9} \rm photons~cm^{-2}~s^{-1}$ \tablenotemark{l} & 0.1 s to 100 s & 2.4 sr \\ 
 \textit{Fermi}/GBM\tablenotemark{c} &  8 keV--40 MeV & 4132 & $0.74~\rm photons~cm^{-2}~s^{-1}$ & 16 ms to 8.192 s & $>8$ sr \\ 
 SWIFT/BAT\tablenotemark{d} &15--150 keV & 5344 & $\sim 10^{-8} \rm ~erg~cm^{-2}~s^{-1}$ & 4 ms to 32 s  & 1.4 sr \\ 
 MAXI/GSC\tablenotemark{e} & 2--30 keV & 3729 & $\sim 7\times 10^{-10} \rm ~erg~cm^{-2}~s^{-1}$ & 1 s to 30 s & $1.5^{\rm o}\times 160^{\rm o}$ \\ 
 \textit{Gaia}\tablenotemark{f} & 330--1050 nm & 2112 & G=20.6~mag (Vega) & 4.5 s & $0.85^{\rm o} \times 0.66^{\rm o}$ \\ 
  PTF/iPTF\tablenotemark{g} & $\approx$ 400--700 nm & 973 & $R\approx 20.6$ (AB) & 60 s & $\approx 8$ sq. deg. \\ 
Pi of the Sky\tablenotemark{h} & $\approx$ 320--900 nm & 756 & V=12~mag (assume Vega) & 10 s & 6400 sq. deg. \\
MMT-9\tablenotemark{i} & $\approx$ 400--800 nm & 644 & V=11~mag (assume Vega) & 0.128 s &  900 sq. deg. \\
Evryscope\tablenotemark{j} & $\approx$ 400--700 nm & 533 & V=16.4~mag (assume Vega) & 120 s & 8660 sq. deg.  \\
\enddata

\tablenotetext{a}{\citet{2016APh....72...76A}.  Major Atmospheric Gamma Imaging Cherenkov telescopes (MAGIC) consists of two Imaging Atmospheric Cherenkov telescopes.  One has been operating since April 2005, the other since 2009 Fall, and both were upgraded in 2012 summer.  Here we calculate the duration from September 2009 and assume 8-hour observation per day.} 
\tablenotetext{b}{\cite{2009ApJ...697.1071A} and https://fermi.gsfc.nasa.gov. The Fermi Gamma-Ray Space Telescope (FGST), was launched on June 11, 2008.}
\tablenotetext{c}{\cite{2009ApJ...702..791M} and https://fermi.gsfc.nasa.gov.  The threshold is for a pulse in the band of 50--300 keV and 1 s peak.}
\tablenotetext{d}{\cite{2005SSRv..120..143B}. The BAT monitor archive begins on February 12, 2005.  The threshold is for a $\sim$ 1 s peak.}
\tablenotetext{e}{\cite{2009PASJ...61..999M}.  MAXI was launched on 2009 July 16.  The threshold corresponds to one International Space Station orbit, in which an objects stay in the FOV for at least 45 s.  For the timescales, we ignore those $\geq$ one scan as they are too long for a ms-scale transient.}
\tablenotetext{f}{\cite{prusti2016gaia};  https://www.cosmos.esa.int/web/gaia; and  http://www.astro.utu.fi/{\textasciitilde}cflynn/galdyn/lecture10.html. The spacecraft was launched on 19 December 2013.}
\tablenotetext{g}{\cite{2009PASP..121.1395L}. Operating from March 2009 to February 2017.}
\tablenotetext{h}{\cite{2014RMxAC..45....7M} and \cite{2014SPIE.9290E..0TC}. The full system started to operate in July 2013. Assume the Vega magnitude system.}
\tablenotetext{i}{\citet{2015BaltA..24..100B}; Mini-MegaTORTORA.  The high time resolution started to operate in June 2014.  Assume the Vega magnitude system.}
\tablenotetext{j}{\cite{2015PASP..127..234L}. The Evryscope-South started to operate since May 2015. We do not include Evryscope-North since it started operations in 2019.  Assume the Vega magnitude system.}
\tablenotetext{k}{The sensitivity corresponds to a 50-hour observation of a point source with Crab-nebula-like spectrum  above 104 GeV.  However, we increase the threshold to $2.4\times10^{-11} ~\rm photons~cm^{-2}~s^{-1}$ to satisfy the requirements that at least one photon is received within 10 ms by the \textit{MAGIC} effective collecting area of $10^9~\rm cm^2$.}
\tablenotetext{l}{The detection threshold corresponds to a one-year survey at high latitude and $>100$ MeV, assuming a source photon spectral index of $-2$.  However, for the lowest timescale of 0.1 s, we increase the threshold to $7\times10^{-8} ~\rm photons~cm^{-2}~s^{-1}$ to satisfy the requirements that at least one photon is received within 0.1 s by the \textit{Fermi}/LAT effective collecting area of $8000~\rm cm^2$.}
\end{deluxetable*}


We demonstrate the application of the methods outlined above using existing high time resolution transient surveys from the near-infrared (NIR) band up to the TeV-band. Following model predictions (see Section~\ref{sec:discussion}), we assume that FRB counterparts are fast transient events shorter than $\sim$ a few minutes. In this work, we only focus on surveys with short cadences ($\lesssim$ 2 minutes), large fields of view (FOVs), and relatively high sensitivities. 

Table \ref{ta:ins} lists the survey instruments  considered in this work.  The survey durations are counted until October 1, 2019.  We assume full-time operation since the launch date for space missions, and a typical average observation time of 8 hours per day since the operation date for ground-based instruments.  We adopt the detection threshold used by each instrument, although some of them correspond to different statistical SNRs, as each survey could have different false-positive rates.  We list the threshold corresponding to the given timescale, unless specified otherwise.  We increase the detection threshold of \textit{MAGIC} and \text{Fermi}/ LAT (at the lower timescale) to $2.4 \times 10^{-11}$ and $7 \times 10^{-8} \rm ~photon~cm^{-2}~s^{-1}$, respectively, to satisfy the requirement that at least one photon is received within the timescales by the corresponding telescopes (step \ref{itm:step_F0}). 

The timescales are chosen differently for the high energy and the optical bands.  All of the high-energy surveys selected in this work have been designed to be sensitive to GRB-like transient events ($\sim$0.1 s to $\sim$100 s).  Each survey has its own transient-candidate self-trigger algorithm that runs on board commensally with observations using a range of trial trigger timescales.  In addition, it is also possible to manually search the survey data afterwards for candidate events using different algorithms and timescales.  For \textit{Fermi}/LAT, we adopt the timescales optimized for FRB-counterpart searching \citep{2019ApJ...879...40C}, since the on-board trigger only responds to very bright bursts due to the high cosmic ray rate.\footnote{Personal communication from Dr. Nicola Omodei.}  For the other high-energy surveys, we list the trial timescales used by the corresponding self-trigger algorithms.  In the optical band, we use the nominal time resolutions for all instruments. 

\section{Results}\label{sec:results}

We estimate $\eta$ for each survey/instrument (Table~\ref{ta:ins}) following the steps introduced in Section~\ref{sec:methods}.  Table~\ref{ta:results} and Fig.~\ref{fig:eta_compare} summarize the results.  In Section~\ref{subsec:results_non_detection}, we make the assumption that no counterpart has been detected to demonstrate the power of our technique.  In Section~\ref{subsec:results_detection}, we investigate the implications of assuming that counterparts exist among the unclassified transient events in some surveys.  

\subsection{Band-to-radio fluence ratios assuming non-detections}\label{subsec:results_non_detection}

\begin{figure*}
\gridline{\fig{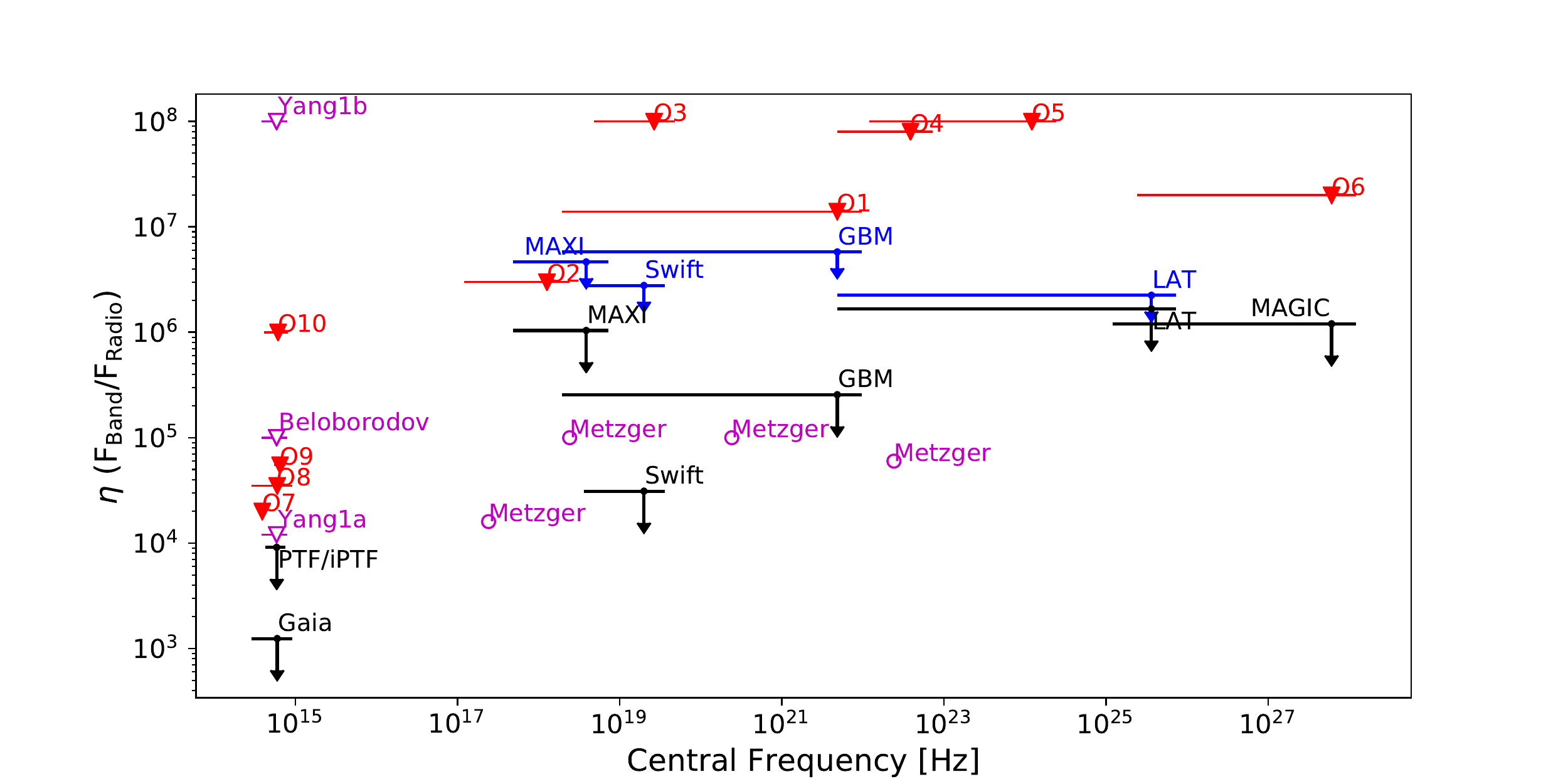}{\textwidth}{(a) Our 95\% confidence upper limits assuming non-detection using the lowest timescales (solid black lines) and the highest timescales  (solid blue lines).}}
\gridline{\fig{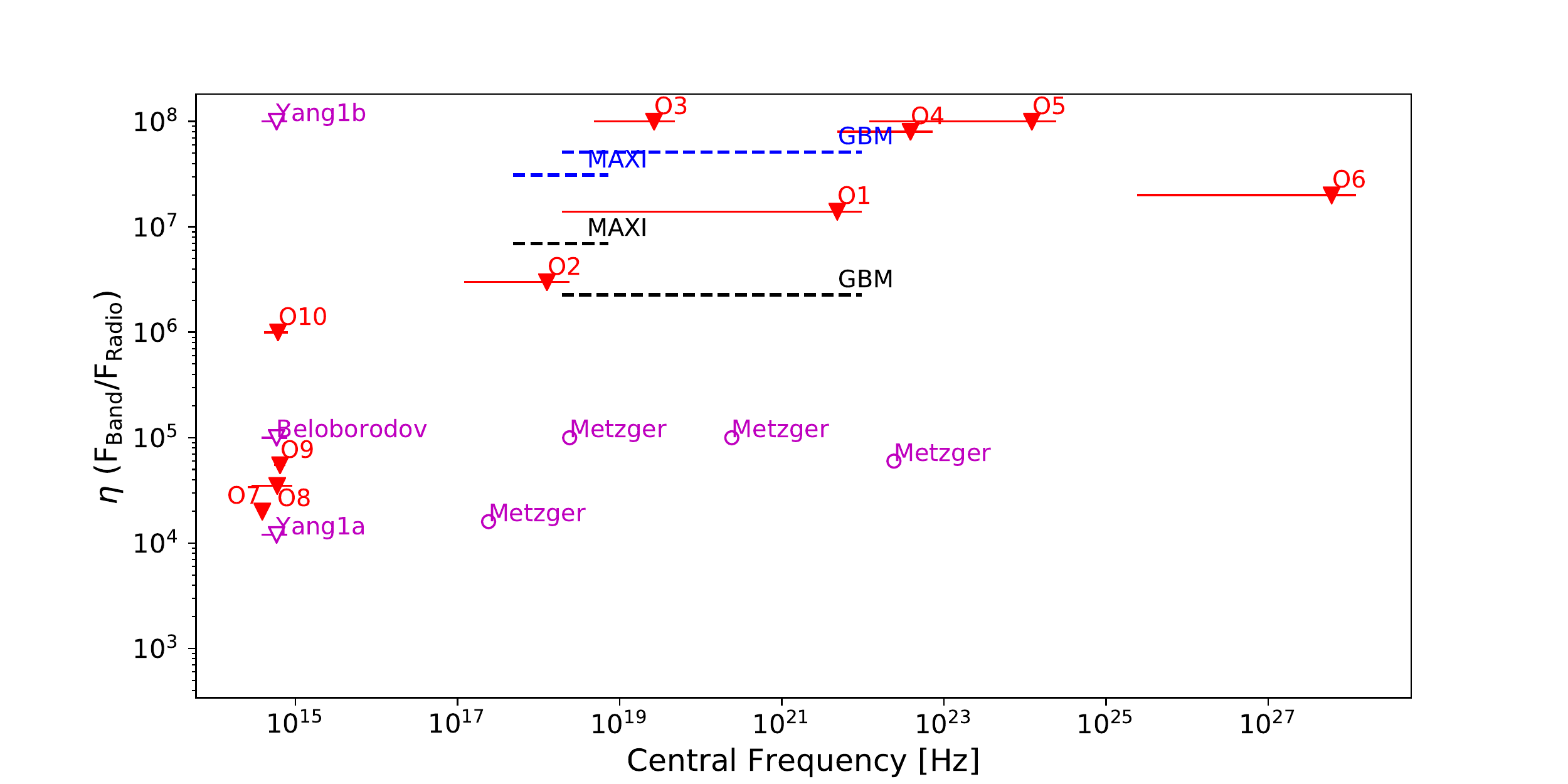}{\textwidth}{(b) Our nominal upper limits assuming that all of the unclassified events in the trigger catalogs were FRB counterparts, using the lowest (dashed black lines) and the highest (dashed blue lines) timescales.}}
\caption{Constraints on the band-integrated fluence ratios from our results (black and blue lines; see section \ref{sec:results}), previous model predictions (the hollow magenta markers are circles for predicted values, and triangles for predicted upper limits; see section \ref{subsec:discussion_models}) and previous observations (filled red triangles; `O1' to `O10' each represents the upper limit from Scholz (`O1' and `O2'), Anumurlapudi, Yamasaki, Casentini, MAGIC Collaboration, Hardy, Wevers, Andreoni and Richmond, respectively; see section \ref{subsec:discussion_obs})}
\label{fig:eta_compare}
\end{figure*}

In Table~\ref{ta:results}, we list 95\% confidence upper limits on the rate of FRB counterparts and on $\eta$ for each survey. We use a reference frequency of $\nu_{\rm 1, Radio} = 1.182 \rm~GHz$ (the lower limit of the Parkes radio band) to convert the specific fluence into the band-integrated radio fluence (in step \ref{itm:step_ratios}). We assume that the counterpart duration $t_{\rm burst}$ is shorter than the timescale $\Delta t$ for all instruments (in step \ref{itm:step_F0} above), and discuss the alternative case in Section~\ref{sec:discussion}. For surveys with multiple timescales, we scale the flux and fluence following steps \ref{itm:step_scale} and \ref{itm:step_F0} using the lowest and the highest timescales.  We list the corresponding results in two rows in Table \ref{ta:results}, and plot both ratios in Fig. \ref{fig:eta_compare} (a). 

Our results are robust within order of magnitude to a selection of variations in the fluence distribution model in Eqn. \ref{eqn:bpl} \citep{2018MNRAS.480.4211M}.  We vary the broken power-law indices $\alpha_1$ and $\alpha_2$ by $\pm0.7$ and find that the results change by $46\%$ ($\alpha_1=-0.48$), $86\%$ ($\alpha_1=-1.88$), $92\%$ ($\alpha_2=-1.5$), and $270\%$ ($\alpha_2=-2.9$), respectively.  We also use a single power-law fluence distribution model with an index of $-1.5$, and find that the results change by less than $88\%$.  However, our results are sensitive to the choice of the photon index in the $\gamma$-ray band.  In step 
\ref{itm:step_f0}, we assume a Crab-like photon index of 2 to calculate the energy flux limit from the photon flux limit for \textit{MAGIC}, \textit{Fermi}/ LAT and GBM.  We vary the photon index to $-2.5$ and $-1.5$ and find that the resulting fluence ratios decrease / increase by 70\% and one order-of-magnitude, respectively. There is little theoretical guidance on what range of photon indices is reasonable for FRB counterparts, but the example of GRBs suggests that photon indices $>-2$ are expected below peak energies (in $\nu F_{\nu}$ spectra) of typically 100\,keV -- 1\,MeV, and photon indices $<-2$ are expected above the peak energies \citep{1998ApJ...506L..23P}.


\begin{deluxetable*}{c c c c c c c}
\tablecaption{Results
\label{ta:results}}

 \tablehead{
  \colhead{Instrument} & \colhead{$\nu_c$}  & \colhead{Flux threshold $f_0$} & \colhead{Fluence threshold $F_0$} & \colhead{Rate\tablenotemark{a}}  & \colhead{$\eta_{\nu}$} & \colhead{$\eta$\tablenotemark{a}} \\
 & \colhead{(Hz)} & \colhead{($\rm erg~cm^{-2}~s^{-1}$)} & \colhead{($\rm Jy~ms~Hz$)} & \colhead{($\rm sky^{-1}~d^{-1}$)} & 
 }
 
 \startdata
 MAGIC & $6.05\times10^{27}$ & $5.54\times10^{-8}$ & $5.54\times10^{16}$ & 21 & $1.18\times10^{-10}$ & $1.20\times10^{6}$ \\ 
 \textit{Fermi}/LAT & $3.63\times10^{25}$ & $3.85\times10^{-7}$ & $3.85\times10^{18}$ & $3.80\times10^{-3}$ & $4.07\times10^{-7}$ & $1.67\times10^{6}$ \\
 & & $5.19\times10^{-10}$ &  $5.19\times10^{18}$ &  &  $5.49\times10^{-7}$ & $2.25\times10^{6}$ \\
 \textit{Fermi}/GBM & $4.84\times10^{21}$ & $6.39\times10^{-7}$  &  $1.02\times10^{18}$ & $1.14\times10^{-3}$ (0.14) & $1.56\times10^{-4}$ & $2.56\times10^{5}$ ($2.27\times10^6$) \\
 & &  $2.82\times10^{-8}$ & $2.31\times10^{19}$ &  &  $3.54\times10^{-3}$ & $5.79\times10^{6}$ ($5.15\times10^7$)\\  
 Swift/BAT & $1.99\times10^{19}$ & $1.58\times10^{-7}$ & $6.32\times10^{16}$ & $5.04\times10^{-3}$ & $1.01\times10^{-5}$ & $3.11\times10^{4}$ \\
  & & $1.77\times10^{-9}$ &  $5.66\times10^{18}$ & & $9.07\times10^{-4}$ & $2.78\times10^{6}$ \\
 MAXI/GSC & $3.87\times10^{18}$ & $4.70\times10^{-9}$ & $4.70\times10^{17}$ & $1.38\times10^{-1}$ (9.00) & $2.54\times10^{-3}$ & $1.04\times10^{6}$ ($6.95\times10^6$) \\
 & & $7.00\times10^{-10}$ &  $2.10\times10^{18}$ & &  $1.14\times10^{-2}$ & $4.65\times10^{6}$ ($3.11\times10^7$) \\
 \textit{Gaia} & $5.97\times10^{14}$ & $6.13\times10^{-14}$ & $2.76\times10^{13}$ & $1.04\times10^{2}$ & $5.15\times10^{-3}$ & $1.24\times10^{3}$ \\
  PTF/iPTF & $5.89\times10^{14}$ & $7.92\times10^{-14}$ & $4.77\times10^{14}$ & $15.9$ & $2.52\times10^{-2}$ & $9.13\times10^{3}$ \\ 
 Pi of the Sky & $6.35\times10^{14}$ & $2.00\times10^{-10}$ & $2.00\times10^{17}$ & $2.56\times10^{-2}$ & $7.30\times10^{-1}$ & $1.06\times10^{5}$ \\
 MMT-9 & $5.62\times10^{14}$ & $5.65\times10^{-10}$ & $7.23\times10^{15}$ & $2.13\times10^{-1}$ & $6.16\times10^{-2}$ & $1.95\times10^{4}$ \\
 Evryscope & $5.89\times10^{14}$ & $4.47\times10^{-12}$ & $5.36\times10^{16}$ & $2.68\times10^{-2}$ & $1.55\times10^{-1}$ & $5.63\times10^{4}$ \\ 
\enddata

\tablenotetext{a}{Rate and $\eta$ results outside parentheses assume no FRB-counterpart detection (Section \ref{subsec:results_non_detection}), and are to be interpreted as 95\%-confidence upper limits. Results in parentheses assume that all unclassified short transients were FRB counterparts (Section \ref{subsec:results_detection}), and are thus to be interpreted as nominal upper limits.}
\end{deluxetable*}

\subsection{What if FRB counterparts have been detected?}\label{subsec:results_detection} 

No compelling candidate FRB counterpart at any wavelength has been reported until the end of 2019.  However, we cannot exclude the possibility that some unclassified short transient events found in existing surveys could be associated with FRBs.  It is beyond the scope of this paper to estimate what fraction of them might be FRB counterparts, but we will investigate the results assuming the extreme case where all of them are FRB counterparts.  We show the results in Table \ref{ta:results} (within parentheses) and Fig. \ref{fig:eta_compare} (b) (dashed lines).  


In the \textit{Fermi}/GBM trigger catalog \footnote{https://heasarc.gsfc.nasa.gov/db-perl/W3Browse/w3hdprods.pl} to the end of 2019,  there were 7045 triggered events, and 370 of them were marked as ``uncertain classification''.  We discuss two limiting cases for these unclassified short transient events.  First, if none of them were FRB counterparts, the results would be the same as those of Section~\ref{subsec:results_non_detection}.  Second, if all of them were FRB counterparts, we estimate $\eta$ using the method as in Section~\ref{sec:methods}.  Using the lowest trigger timescale as an example, steps 1 to 3 remain the same, so the \textit{Fermi}/GBM fluence limit is still $1.02\times10^{18}$ Jy ms Hz.  In step \ref{itm:step_rate}, the rate is now estimated as $R = \frac{370}{4132~\rm days}\frac{4\pi~\rm sky^{-1}}{8} = 0.14~\rm day^{-1} sky^{-1}$, $\sim$ 100 times larger than the upper limit assuming non-detection.  In the radio band, the fluence threshold that would have produced the new detection rate is 380 Jy ms, or $4.49\times10^{11}$ Jy ms Hz.  Hence, $\eta=\frac{1.02\times10^{18}}{4.49\times10^{11}}\approx2.27\times10^6$, $\sim~10$ times higher than our result in Section \ref{subsec:results_non_detection}.  


In the \textit{MAXI} trigger catalog\footnote{https://gcn.gsfc.nasa.gov/maxi\_grbs.html} between 2011 April 18 and 2020 January 28, 168 events were classified as ``either GRB or unknown X-ray transient''.  If all of them were FRB counterparts, the rate would be $R = \frac{168}{3207~\rm days} \frac{4\pi~\rm sky^{-1}}{0.0731} = 9.00~\rm day^{-1} sky^{-1}$. The radio fluence that could have produced the same rate would be $6.76\times10^{10}$ Jy ms Hz, and the fluence ratio would be $\eta=\frac{4.70\times10^{17}}{6.76\times10^{10}}\approx6.95\times10^6$, $\sim~7$ times higher than our previous result. 

In the optical band, we only use the most constraining result (from \textit{Gaia}) in the following comparison with model predictions and previous observations.  No unclassified fast \textit{Gaia} transient has been reported\footnote{\citealt{2018MNRAS.473.3854W} develop a method to search the \textit{Gaia} data for fast transients between tens of seconds to hours.  They find four events produced by stellar flares in 23.5 square degrees of sky.}, and so we tentatively maintain the non-detection assumption for \textit{Gaia} in our results. This may change as more \textit{Gaia} data are searched for unclassified fast transients.


\section{Discussion}\label{sec:discussion}

We have developed and demonstrated a technique to estimate the ratios between FRB energy output in the radio band, and in various bands from the near-IR to $\gamma$-rays. Preliminary results based on published surveys for fast transients (Table~\ref{ta:ins}) are presented in Table~\ref{ta:results} and Fig.~\ref{fig:eta_compare} in two extreme cases: assuming either that no FRB counterpart has been detected, or assuming that all unclassified fast transients are FRB counterparts. Our method uses the statistical properties of the observed FRB population, while most of the previous observational estimations are based on multi-wavelength observations at the locations of individual FRB events.  In this section, we will compare our results with theoretical predictions and previous observations, and briefly discuss possible future FRB counterpart search strategies. 

\subsection{Comparison to theoretical Predictions} 
\label{subsec:discussion_models}

The two leading FRB emission mechanisms are the synchrotron maser and coherent curvature radiation.  We first compare our high-energy results with predictions from these two types of models, as well as a class of models generally involving contemporaneous high-energy flares, and the cosmic comb model.  We then compare our optical results with model predictions made under five scenarios.  Finally, we conclude by calculating the model-predicted counterpart detection rate for some of the existing surveys (Table~\ref{ta:rate}). This section is partly intended as a pedagogical resource for future studies of FRB multi-wavelength counterparts.

\begin{deluxetable*}{ccccccccccc}
\rotate
\tablecaption{ Theoretical Predictions and Expected Counterpart Rate
\label{ta:rate}}

 \tablehead{
  \colhead{Model} & \colhead{Band} & \colhead{$t_{\rm Counterpart}$} & \colhead{$\eta_{\nu}$} & \colhead{$\eta$} & \colhead{Instrument} & \colhead{Timescale\tablenotemark{a}} & \colhead{Rate} & \colhead{$\langle n\rangle$ \tablenotemark{b}} & \colhead{Survey Duration \tablenotemark{c}}\\
  & & & & & & & \colhead{(sky$^{-1}$ day$^{-1}$)} & \colhead{(days)} & \colhead{(days)} 
 }
 
 \startdata
  Metzger & $\nu_1$=100 MeV & $\sim$ 3 ms & $\sim~2.8\times10^{-9}$ & $\sim 6.0 \times 10^4$ & \textit{Fermi}/LAT & 0.1 s & $2.29 \times 10^{-6}$ & $2.29 \times 10^6$ & 4132\\
  Metzger & $\nu_1$=1 MeV & $\sim$ 50 ms & $\sim~4.1\times10^{-7}$ & $\sim 10^5$ & \textit{Fermi}/GBM & 64 ms & $2.13\times10^{-5}$ & $7.36\times10^4$ & 4132\\
  Metzger & $\nu_1$=100 keV\tablenotemark{d}& $\sim$ 1 s & $\sim ~4.1\times10^{-6}$ & $\sim 10^5$ & \textit{Swift}/ BAT & 1.024 s & $10^{-4}$ & $8.95\times 10^4$ & 5344 \\ 
  Metzger & $\nu_1$=10 keV & $\sim$ 1 s & $\sim ~4.1\times10^{-5}$ & $\sim 10^5$ & \textit{Fermi}/GBM & 1.024 s &$1.01\times10^{-6}$ & $1.55\times10^6$ & 4132\\ 
  Beloborodov & Optical & $\sim$ 1 s & $\lesssim 2.0\times 10^{-1}$ & $\lesssim 10^5$ & \textit{Gaia} & 4.5 s & $\lesssim 6.31 \times 10^5$ & $\gtrsim 1.17 \times 10^{-1}$ & 2112\\ 
  Yang 1a\tablenotemark{e} & Optical & $\sim$ 1 ms &  $\lesssim 2.0\times 10^{-2}$ & $\lesssim 1.2\times 10^4$ & \textit{Gaia} & 4.5 s & $\lesssim 3.98\times10^3$ & $\gtrsim 18$ & 2112\\
  & & & $\gtrsim 1.7\times 10^{-13}$ & $\gtrsim 1.2\times 10^{-7}$ &\textit{Gaia} & 4.5 s & $\gtrsim 1.76\times10^{-21}$ & $\lesssim 4.18\times10^{25}$\\ 
  Yang 1b & Optical & $\sim$ a few $\times 10$ s & $\lesssim 1.7\times 10^2$ & $\lesssim 10^8$ & \textit{Gaia} & $\approx$45 s & $\lesssim 1.40\times10^{11}$ & $\gtrsim 5.27\times10^{-7}$ & 2112\\
  Yang 1c & Optical & $\sim$ 1 ms & $\lesssim 6.6\times10^{-7}$ & $\lesssim 3.8\times10^{-1}$ & \textit{Gaia} & 4.5 s & $\lesssim 5.51\times10^{-7}$ & $\gtrsim 1.34\times10^{11}$ & 2112\\
  Yang 2a & Optical & $\sim$ 1 ms & $\lesssim 6.6\times10^{-11}$ & $\lesssim 3.8 \times 10^{-5}$ & \textit{Gaia} & 4.5 s & $\lesssim 8.73\times10^{-16}$ & $\gtrsim 8.43\times10^{19}$ & 2112\\
  Yang 2b & Optical & $\sim$ 1 ms & $\lesssim 6.6\times10^{-10}$ & $\lesssim 3.8 \times 10^{-4}$ &\textit{Gaia} &4.5 s & $\lesssim 1.38\times10^{-13}$ & $\gtrsim 5.32\times10^{17}$ & 2112\\
 \enddata
\tablenotetext{a}{The lowest instrumental timescale above the theoretical counterpart duration.}
\tablenotetext{b}{The expected number of days to detect one single counterpart using the corresponding instrument (Poisson errors neglected).} 
\tablenotetext{c}{Same as the effective duration in Table \ref{ta:ins}.}
\tablenotetext{d}{$\eta$ interpolated between 1 MeV and 10 keV.}
\tablenotetext{e}{e.g. ``1a'' refers to case 1, scenario (a) in the model (Section \ref{subsec:discussion_models}).} 

\end{deluxetable*}

\subsubsection{\cite{2019MNRAS.485.4091M} model}

Synchrotron masers have been widely discussed as an astrophysical coherent emission process (e.g. \citealt{1991PhFlB...3..818H}; \citealt{2018ApJ...864L..12L}), and one common variation is coherent emission from synchrotron masers produced by ultrarelativistic shock in magnetized plasmas (e.g. \citealt{1988PhRvL..61..779L}; \citealt{2014MNRAS.442L...9L}; \citealt{2017ApJ...843L..26B}; \citealt{2019arXiv190807743B};  \citealt{2019MNRAS.485.4091M}; \citealt{2019arXiv191105765M}).  \cite{2019MNRAS.485.4091M} describe a model using the particle-in-cell (PIC) simulation results for maser emission and the dynamics of self-similar shock deceleration.  Magnetar flares eject ultra-relativistic ion-electron shells that are supersonic into the surrounding magnetized trans-relativistic ion-electron plasma released by previous flares.  The forward shock creates a population inversion and enables the synchrotron maser process, which results in narrowly peaked coherent radio emission putatively responsible for FRBs.  The same forward shock, however, primarily dissipates energy through a synchrotron ``afterglow'' that will result in observable high-energy counterparts.  The model predicts that the observed counterpart luminosity is $L_{\rm \gamma} \sim 10^{45}$-- $10^{46}~\rm erg~s^{-1}$ with a duration of $\sim 0.1-10$ ms in the MeV--GeV band, and $L_{\rm X} \sim 10^{42}-10^{53}~\rm erg~s^{-1}$ with a duration of $\sim0.1$--1 s in the keV band. A weak optical counterpart is possible if the upstream plasma were composed of electrons and positrons rather than electrons and ions; we do not consider any resulting quantitative predictions here.  We convert the above predictions to fluence ratios $\eta$, which can be directly compared with our results.  Assuming a typical FRB of duration 1\,ms initiated by a flare of energy $\sim 10^{44}$\,erg, the results of \cite{2019MNRAS.485.4091M} imply ratios of $6 \times 10^4$, $10^5$, $10^5$ and $2 \times 10^4$ for a counterpart whose band starts from 100 MeV, 1 MeV, 10 keV and 1 keV, respectively (hollow magenta circles in Fig \ref{fig:eta_compare}). 

\subsubsection{\cite{2019arXiv190807743B} model}

Meanwhile, \cite{2019arXiv190807743B} proposes that the synchrotron maser is formed instead when the magnetar giant flares launch ultra-relativistic blast waves ($\Gamma \gtrsim 10^3$) into the relativistic ($\Gamma \sim 10^2$), persistent  magnetar wind outflow, which consists of $e^{\pm}$.  A bright optical counterpart occurs only when the blast wave strikes a hot wind bubble in the slow ion tail of a previous flare. The optical flash is estimated to have a duration of $\sim 1$ s and an energy upper limit of $\sim 10^{44}$ erg.  The optical-to-radio fluence ratio would be $\eta \lesssim 10^5$ using the average FRB 121102 burst energy of $10^{39}$ erg \citep{2017ApJ...850...76L}, and $\eta \sim 10^3$ using the FRB energy corresponding to the strongest explosion, which produces the brightest optical flash in their model.  The former is shown in Fig. \ref{fig:eta_compare} to compare with our results, since our technique utilizes the statistical features of the entire FRB population.  The latter prediction could be comparable to results of simultaneous multi-wavelength observations of individual events.  Note that in this model  many FRBs do not have optical counterparts, since only strong magnetar flares may have significant ion tails. 

\subsubsection{Soft gamma-ray repeater (SGR) giant flares as FRB counterparts}
\label{subsubsec:sgrs}

More generally, in many FRB models the emission processes are initiated by SGR giant flares. The energy released by giant flares is typically specified in the $\gamma$-ray band, where the \textit{Fermi}/GBM survey suggests constraints ranging between $\eta\lesssim10^{5}$ (0.1-s counterparts, assuming no extant detections) and $\eta\sim10^{7}$ (100-s counterparts, assuming all unclassified events are giant flares). These constraints can be used to test the hypothesis that each giant flare corresponds to an FRB. 

Following \citet{2007ApJ...659..339O}, the rate of giant flares in the Milky Way, which hosts four SGRs, is $\lesssim0.002$\,yr$^{-1}$ for energies $E_{\rm SGR}>4\times10^{46}$\,erg (this rate is based on an analysis of extragalactic giant-flare candidates), and $\sim0.1$\,yr$^{-1}$ for energies $E_{\rm SGR}>2\times10^{44}$\,erg (Poisson errors neglected; this rate is derived from the Milky Way alone). \citet{2007ApJ...659..339O} derives the number of SGRs in a given galaxy by comparing its core-collapse supernova rate with that of the Milky Way, which is justified given the short lifetimes ($O({\rm kyr})$) of active SGRs. As the overall star-formation rate is a reasonable proxy for the core-collapse supernova rate \citep{2014ARA&A..52..415M}, we can derive the (local) volumetric rate of giant flares by scaling the Milky Way rate by the ratio of the local star-formation rate density \citep[ $0.015\,M_{\odot}$\,yr$^{-1}$\,Mpc$^{-3}$;][]{2014ARA&A..52..415M} and the Milky Way star-formation rate \citep[$1.9\,M_{\odot}$\,yr$^{-1}$;][]{2011AJ....142..197C}. For $E_{\rm SGR}>4\times10^{46}$\,erg and $E_{\rm SGR}>2\times10^{44}$\,erg, the volumetric giant flare rates are $\lesssim2\times10^{4}$\,Gpc$^{-3}$\,yr$^{-1}$ and $\sim8\times10^{5}$\,Gpc$^{-3}$\,yr$^{-1}$ respectively. If each giant flare produces an FRB, these volumetric rates correspond to estimates of the FRB volumetric rates for radio-band energy releases $E_{\rm FRB} = E_{\rm SGR}/\eta$ (e.g., $E_{\rm FRB}=4\times10^{39} \rm~erg$ and $E_{\rm FRB}=2\times10^{37}\rm~erg$ respectively for $\eta=10^{7}$). The volumetric rate of FRBs in the local Universe inferred from the Canadian Hydrogen Intensity Mapping Experiment (CHIME) is $\sim10^{5}$\,Gpc$^{-3}$\,yr$^{-1}$ \citep{2019NatAs...3..928R}, approximately above an energy threshold of $2\times10^{37}$\,erg\,s$^{-1}$.\footnote{An energy of $E_{\rm FRB}\sim2\times10^{37}$ erg corresponds to a 2\,Jy\,ms burst detected by CHIME at 100\,Mpc.} Thus, contrary to previous studies \citep{2014ApJ...797...70K,2019NatAs...3..928R}, the giant-flare rate may in fact be too high to explain the FRB rate. This result would be strengthened if the value of $\eta$ is substantially lower than the conservative upper bound of $10^{7}$ derived herein. We note, however, that all rate estimates above are subject to severe Poisson errors, and that this analysis will require significant refinement before firm conclusions can be drawn. 

A similar analysis can be applied to any multiwavelength event that triggers an FRB. This is of particular relevance to the ``cosmic comb'' model, where a regular pulsar magnetosphere is ``combed'' by a nearby strong plasma stream with a ram pressure higher than the magnetic pressure in the magnetosphere.  The stream triggers magnetic reconnection that accelerates particles within the magnetosphere, which produce coherent emission by the curvature-radiation or cyclotron instability mechanisms.  On the one hand, when the plasma stream comes from nearby energetic events, such as active galactic nucleus (AGN) flares, those events should be detected as FRB counterparts.  On the other hand, when the stream comes from closer but less luminous events, such as stellar flares from a companion star, no detectable counterpart would occur.

\subsubsection{Curvature radiation}

Another commonly discussed coherent emission process is curvature radiation (e.g. \citealt{2017MNRAS.468.2726K}; \citealt{2018MNRAS.477.2470L}).  For example, \cite{2018MNRAS.477.2470L} propose a model where counter-streaming $e^{\pm}$ plasma inside the twisted magnetosphere of a magnetar rapidly clumps due to the two-stream instability.  When magnetic reconnection occurs near the magnetar surface, these clumps are accelerated along magnetic field lines and radiate coherently.  The model predicts fluence ratios of $\eta \sim 1$ in all bands, so there would be no detectable FRB counterpart.


\subsubsection{Fast optical bursts associated with FRBs}

We compare our most constraining result from \textit{Gaia} ($\eta \lesssim 10^3$, assuming non-detection and using the \textit{Gaia} time resolution of 4.5 s) with predictions made by \citealt{2019ApJ...878...89Y}.  These authors investigate the detectability of the ``fast optical bursts'' (FOBs) associated with FRBs in two broad cases and five specific scenarios.  We convert them to the constraints on the optical-to-radio fluence ratios assuming an FRB of 1-ms duration (hollow magenta triangles in Fig. \ref{fig:eta_compare}).  
\begin{description}

\item[Case 1] FOB formed by inverse Compton scattering between the FRB photons and ambient electrons.
\renewcommand{\theenumi}{\alph{enumi}}
\begin{enumerate}
    \item OB and FRB both formed in the pulsar magnetosphere ($10^{-7} \lesssim \eta \lesssim 10^4$, $t_{\rm FOB} \sim 1$ ms).
    \item FOB formed in a surrounding nebula, and FRB near the neutron star ($\eta \lesssim 10^8$, $t_{\rm FOB} \sim$ a few $\times$ 10 s).
    \item FRB formed by  synchrotron maser mechanism and FOB formed by inverse Compton scattering between the maser electrons and the FRB photons ($\eta \lesssim 0.38$, $t_{\rm FOB} \sim 1$ ms). 
\end{enumerate}

\item[Case 2] FOB and FRB formed by the same emission mechanism.
\renewcommand{\theenumi}{\alph{enumi}}
\begin{enumerate}

\item Curvature radiation by particle bunches ($\eta \lesssim 3.8 \times 10^{-5}$, $t_{\rm FOB} \sim 1$ ms). 

\item Synchrotron maser ($\eta \lesssim 3.8 \times 10^{-4}$, $t_{\rm FOB} \sim 1$ ms).

\end{enumerate}

\end{description}


We omit case 1(c) and case 2 in Fig. \ref{fig:eta_compare} as they are too low to compare with any existing optical telescope. Compared with our \textit{Gaia} result, the upper limits predicted by cases 1(a) and 1(b) are greater by $\sim$ 1 and 5 orders of magnitude, while predictions of the other three scenarios are lower ny $\sim$ 4, 8 and 7 orders of magnitude, respectively.  Hence, a detection of FRB counterparts by \textit{Gaia} (or indeed any other optical telescope) would rule out case 1(c) and case 2, and it might be able to rule out case 1(a) if the observed fluence ratio lies between $\sim~10^4$ and $10^8$.  

\subsubsection{How many multi-wavelength FRB counterparts should blind surveys detect?}

Finally, we estimate the counterpart detection rate from each model (Table \ref{ta:rate}) by combining the model-predicted $\eta$ with the observed FRB fluence distribution.  For example, the theoretical counterpart duration at 1 MeV ($2.4\times10^{20}$ Hz) from the \cite{2019MNRAS.485.4091M} model is $\sim$ 50 ms.  The lowest trigger timescale of \textit{Fermi}/GBM above 50 ms is 64 ms.  At this timescale, the GBM band-integrated fluence threshold is $F_{0,\rm GBM} = 2\times 10^{18}$ Jy ms Hz, and $F_{\nu,0,\rm GBM} \approx \frac{2\times 10^{18} \rm~Jy~ms~Hz}{2.4\times10^{20}\rm~Hz} \approx 8.4 \times 10^{-3}$ Jy ms.  The model predicts that $\eta_{\nu} \approx 4.1\times10^{-7}$ at $\nu_1 = 1$ MeV, so the expected counterpart rate above $F_{\nu,0,\rm GBM}$ in that band would be $\tilde{R}(F_{\nu}) = R(\frac{F_{\nu,0,\rm GBM}}{\eta_{\nu}}) = R(2.1\times10^4 \rm~Jy~ms) \approx 2.13 \times 10^{-5} \rm~sky^{-1}~day^{-1}$ (Equations \ref{eqn:rate} and \ref{eqn:bpl2}).  On average, \textit{Fermi}/GBM is expected to detect one such event per $\langle n \rangle \approx (2.13 \times 10^{-5})^{-1} (\frac{4\pi}{8 \rm~sr}) \approx 7.36 \times 10^4$ days.  By comparing $\langle n \rangle$ with the relevant survey durations $n$ (the last two columns in Table \ref{ta:rate}), it is unlikely that \textit{Fermi}/ LAT or GBM have already detected any counterparts events, or will detect one in the near future, according to the prediction of \citet{2019MNRAS.485.4091M}.  

As another example, we interpolate the \citet{2019MNRAS.485.4091M} predictions and conservatively assume a 1-s duration counterpart in the \textit{Swift}/ BAT band (100 KeV) with $\eta_{\nu} \approx 4.1 \times 10^{-6}$. At this timescale, $F_{\nu,0,BAT} \approx 4.18 \times 10^{-2} \rm~ Jy~ms$, and the expected counterpart rate in that band would be $R(\frac{4.18 \times 10^{-2} \rm~ Jy~ms}{4.1 \times 10^{-6}}) = R(10^4\rm~Jy~ms) \approx 10^{-4} \rm~sky^{-1}~day^{-1}$, and $\langle n \rangle \approx 8.95 \times 10^4$ days.  \textit{Swift}/ BAT is not expected to have detected any FRB counterpart based on this model, unless the counterpart duration at 100 keV is significantly shorter ($\lesssim 64$ ms).

We cannot comment on the predictions from \citet{2017ApJ...843L..26B} and \citet{2019ApJ...878...89Y}, as their models only indicate the lower limits to $\langle n \rangle$.  

\subsubsection{Caveat emptor}

Cautions should be taken in the comparisons described above.  First, our technique relies on a homogeneous FRB population distribution, and the synchrotron-maser and curvature-radiation models may not be able to explain so far non-repeating FRBs.  Although it has been argued that all FRB sources repeat in their lifetimes \citep{2019NatAs...3..928R}, some sources are clearly more active than the others and it is not yet clear whether or not they belong to the same population groups.  Second, the high-energy counterpart could be either the giant flare that initiates the FRB emission processes or the afterglow, or both, but their contributions are observationally indistinguishable.  Third, it might be difficult to distinguish intrinsic emissions from propagation effects, since any dense intervening medium has different attenuation effects on different wavelengths.  For example, in the model of  \citet{2019MNRAS.485.4091M}, it is unclear whether the keV photons would escape from supernova ejecta shells surrounding the proposed magnetars, or get absorbed by the neutral gas on the FRB timescale \citep{2018MNRAS.481.2407M}.  This ambiguity makes it difficult to constrain the model based on the non-detection of X-ray counterparts.  Fourth, in this work we only focus on surveys with cadences less than 2 minutes, but longer-duration counterparts may also be possible \citep[e.g.,][]{2015MNRAS.447..246P}. In this case, sensitive surveys on these longer timescales should also be considered.

\subsection{Comparison to other observations}
\label{subsec:discussion_obs}

As is evident from Figure~\ref{fig:eta_compare}, our technique generally provides stronger constraints on $\eta$ than previous observations. We consider a selection of previous observational results in turn. 



Some high-energy transient surveys have been blindly searched for FRB counterpart candidates without using knowledge of individual FRB events.  \citet{2016MNRAS.460.2875Y} (O4 in Figure~\ref{fig:eta_compare}) performed a blind search for $\gamma$-ray flashes (duration 1 to 10 ms) using the 7-year \textit{Fermi}/LAT data.  No event is found after removing flashes associated with known steady $\gamma$-ray sources and false events produced by the diffuse background.  They found a $\gamma$-ray to radio fluence ratio of $\eta \lesssim (4.2 \sim 12)\times 10^7$ by modeling FRBs as standard candles with a power-law $\gamma$-ray spectrum and estimating the comoving FRB rate density using the nine FRB detected by then.  In comparison, our technique adopts a model-independent FRB population distribution based on a directly measurable quantity (fluence) and a significantly larger sample ($\sim$ 50).  Using our technique, we find $\eta \lesssim 1.7 \times 10^6$ (100 ms) based on the non-detection in the 7-year of \textit{Fermi}/LAT data.

In the optical band, we estimate $\eta$ from a few survey sub-datasets that have been blindly searched for fast transients.  \citet{2018MNRAS.473.3854W} (O8 in Figure~\ref{fig:eta_compare}) develop a method to blindly search the \textit{Gaia} Photometric Science Alerts data base for fast transients between tens of seconds to hours.  They demonstrate the method on a trial data set that spans $\sim$ 23.5 deg$^2$ of sky and has been repeatedly scanned for 40 to 50 times.  Four events produced by stellar flares are found but no unclassified event is detected.  The non-detection implies a optical-to-radio-band fluence ratio of 
$\eta \lesssim 4 \times 10^4$ 
using our technique.  In addition, \citet{2020MNRAS.491.5852A} (O9 in Figure~\ref{fig:eta_compare}) specifically search for extragalactic fast optical transients with durations down to 70 s using the Dark Energy Camera as part of the Deeper Wider Faster programme.  The g-band limiting magnitude of one single exposure (20 s) is $\sim$ 23 mag (AB), the FOV is 2.52 deg$^2$ and the total observation time of their data set is 25.76 hours.  Four events with uncertain classifications are detected, but no $\gamma$-ray signal or FRB is found within $\pm$ 1 day near these transients.  Using our technique, the optical-to-radio-band fluence ratio is 
$\eta \lesssim 6 \times10^4$.  
Finally, \citet{2019arXiv191011343R} (O10 in Figure~\ref{fig:eta_compare}) find no transients with durations from 1.5 s to 11.5 s using the \textit{Tomo-e Gozen} wide-field CMOS mosaic camera data (limiting magnitude V=15.6) with a control time of 177,502 $\rm deg^2~sec$.  The non-detection implies that $\eta \lesssim 10^6$ using our technique. 

Most previous constraints on $\eta$ are based on counterpart searches in the sky region of individual FRB events, either contemporaneous or not.  We summarize them below and show some of the  stronger constraints in Fig. \ref{fig:eta_compare}. 

\citet{2018MNRAS.481.2479M} (O6 in Figure~\ref{fig:eta_compare}) conducted simultaneous observations of the repeating FRB 121102 using the Arecibo telescope and MAGIC (100 GeV--50 TeV, and the optical band).  Five FRBs were detected during this time (mean fluence $\sim$ 2 Jy ms), but no simultaneous or persistent counterparts were found by MAGIC.  This implies that $F_{> 100 \rm GeV} / F_{\rm Arecibo} \lesssim 2\times10^7$ for a 10-ms counterpart, and $F_{\rm optical} / F_{\rm Arecibo} \lesssim 4\times10^3$ for a 1-ms counterpart. 

\cite{2019arXiv191110189C} (O5 in Figure~\ref{fig:eta_compare}) searched the AGILE archival data for MeV--GeV  counterparts of two repeating FRB sources.  They find no prompt emission and estimate a band-integrated fluence ratio of $F_{\rm MeV}/F_{\rm Radio} \lesssim 10^8$, assuming ms-scale emissions in the MeV band. 

\citet{2019ApJ...879...40C} found no prompt high energy counterpart with durations between 0.1 and 100 s for a sample of 23 FRBs in the \textit{Fermi}/GBM, \textit{Fermi}/LAT and \textit{Swift} data.  They estimate the fluence ratio to be $\eta \lesssim 10^{7-12}$ for the timescale of 0.1 s (and $\eta \lesssim 10^{8-13}$ for 100 s).

\citet{2019arXiv191100537A} (O3 in Figure~\ref{fig:eta_compare}) find no prompt X-ray counterpart for a sample of 42 FRBs in the \textit{AstroSat}/CZTI data (20--200 keV) and estimate the fluence ratio to be $F_{\rm X}/F_{\rm Radio} \lesssim 10^{8-10}$.

\cite{2017ApJ...846...80S} (O1 and O2 in Figure~\ref{fig:eta_compare}) present simultaneous observations of the repeating FRB 121102 using the XMM-\textit{Newton}, \textit{Chandra}, and \textit{Fermi}/GBM telescopes along with several radio telescopes.  They found 12 radio bursts and no contemporaneous counterpart emission.  They estimate that $\eta \lesssim 4\times 10^8$ in the \textit{Fermi}/GBM band and $\eta \lesssim 3\times10^6$ in the X-ray band (0.5-- 10 keV) assuming bursts of $< 700$ ms.  In addition, they find no X-ray counterpart in the sky region at anytime during these observations. Using the fluence distribution of radio bursts from FRB 121102, they estimate that $\eta \lesssim 5\times10^7$ in the \textit{XMM-Newton} band (0.1-- 15 keV) and $\eta \lesssim 10^8$ in the Chandra band (0.5-- 7 keV), assuming 5-ms X-ray bursts. 

\cite{2017MNRAS.472.2800H} (O7 in Figure~\ref{fig:eta_compare}) conducted simultaneous observations of  FRB 121102 using the high-speed optical camera ULTRASPEC on the Thai national telescope, and the Effelsberg radio telescope.  They detected 13 radio events and no prompt optical counterparts.  They compared the median radio fluence of those bursts with the optical detection limit and find $F_{\nu, \rm 767 nm}/F_{\nu,\rm 1.4 GHz} \lesssim 0.077$, corresponding to a band-integrated fluence ratio of $\eta \lesssim 2 \times 10^4$. 


Finally, in a class of FRB models, the emission processes are initiated by SGR giant flares.  \citet{2016ApJ...827...59T} estimate that $F_{\nu,\rm 1.4 GHz} / F_{\gamma} \lesssim 10^7 \rm~Jy~ms~erg^{-1}~cm^2$ for a 10-ms radio fluence based on the radio non-detection of a $\gamma$-ray giant flare from the magnetar SGR 1806-20.  
Their results imply $\eta = F_{\gamma} / F_{\rm radio} \gtrsim 10^{10}$, which is inconsistent with our $\gamma$-ray upper limits and thus in tension with the idea that SGR flares generally produce FRBs. This outcome is consistent with the discussion in Section~\ref{subsubsec:sgrs}.  


 \subsection{Future searching strategies}
 \label{subsec:discussion_strategies}
 
In the absence of FRB counterpart detections, $\eta$ can be constrained from either simultaneous multi-wavelength and radio searches, or blind searches combined with the FRB population (considered in this paper).  The observational strategies are slightly different in these two cases, given the nature of the FRB fluence distribution. 

In a simultaneous counterpart search with non-detection, the upper limit to $\eta \propto f_0 \cdot \Delta t$. Here, $\Delta t$ is the timescale and $f_0$ is the corresponding detection flux threshold.  The FOV ($\Omega$) makes no difference.
The total observation duration ($L$) is also irrelevant, until the next FRB occurs. 

In contrast, in a blind search where no counterpart is found, the upper limit on $\eta$ also depends on $L$ and $\Omega$.  The counterpart rate upper bound $R \propto L^{-1} \cdot \Omega^{-1}$ (step \ref{itm:step_rate} in section \ref{sec:methods}).  Using the broken power law fluence distribution, the radio fluence threshold that would have produced the same rate $R$ is $F_{\rm \nu,0,Radio} \propto R^{1/\alpha} \propto L^{-1/\alpha} \cdot \Omega^{-1/\alpha}$.  Meanwhile, the fluence  threshold in the band of the counterpart is $F_{\rm \nu,0,Band} \propto f_0 \cdot \Delta t$.  Therefore, 
\begin{equation} 
\begin{split} 
\label{eqn:eta_propto}
\eta & \propto f_0 \cdot \Delta t \cdot \Omega^{1/\alpha} \cdot L^{1/\alpha}, \\
&\propto 10^{\rm-0.4~m} \cdot \Delta t \cdot \Omega^{1/\alpha} \cdot L^{1/\alpha}.
\end{split} 
\end{equation}
Here, $m$ is the absolute magnitude, and $\alpha$ is the power-law index of the fluence distribution (Eqn. \ref{eqn:bpl}).  Assuming non-detection or low counterpart detection rate, the power-law index of the ASKAP FRB sample ($\alpha_2 = -2.2$) is more relevant than that of the Parkes ($\alpha_1 = -1.18$), since the former describes events with rate below $R_b \approx 170 \rm~sky^{-1}~day^{-1}$ (Eqn. \ref{eqn:bpl2}), and the latter describes more common events (Eqn. \ref{eqn:bpl1}).  Using $\alpha_2$, $\eta \propto f_0 \cdot \Delta t \cdot \Omega^{-0.45} \cdot L^{-0.45}$, so one could enhance the constraint on $\eta$ by one order of magnitude by lowering the detection flux threshold by 10 times, using a 10 times shorter timescale, or increasing the FOV or survey duration by 158 times (15 times if using $\alpha_1$).  


Nonetheless, there may be reasons to also require large fields of view or survey areas.  If counterpart events are rare in the local Universe, a significant number of nearby galaxies would need to be included in a blind search. Our technique assumes that FRBs are uniformly distributed over the sky region scanned by a survey, which is likely true for distant FRBs, but may not be true for nearby FRBs or near the Galactic plane.

\section{Conclusions}
\label{sec:conclusion}


We have developed and demonstrated a technique to estimate $\eta$ -- the ratio between the energy emitted by the multi-wavelength counterparts of FRBs and FRBs themselves -- by combining existing multi-wavelength fast transient surveys with the fluence distribution of the FRB population. The extremely large fields of view and observation durations of surveys from the optical to the TeV bands, combined with the high all-sky rate of FRBs, mean that the locations of several FRBs undetected by radio telescopes have likely been observed by telescopes across the electromagnetic spectrum. We use the properties of several multi-wavelength surveys (listed in Table~\ref{ta:ins}) to constrain $\eta$ under the assumption that no FRB counterparts have been detected, and in some cases to estimate $\eta$ under the assumption that all unclassified transient events are FRB counterparts (Table~\ref{ta:results} and Figure~\ref{fig:eta_compare}). We conclude the following:
\begin{enumerate}
    
    \item Even our most conservative constraints / estimates for $\eta$ are lower than several existing results, which are largely based on targeted observations of known FRB locations, coordinated between multiple telescopes.

\item The FRB models proposed by \citet{2019MNRAS.485.4091M} and \citet{2019arXiv190807743B}, which involve synchrotron masers initiated by shocks driven by young-magnetar flares, are closest to our constraints on $\eta$. In some scenarios, \textit{Gaia} should have already detected several FRB counterparts. FRB counterparts may be found amongst unclassified transient events. This demonstrates the power of our technique to address FRB model predictions. However, in the high-energy bands, surveys by the \textit{Fermi} and \textit{Swift} satellites are not likely to have detected FRB counterparts unless the photon indices are significantly steeper than $-2$.

    \item Our technique can also be used to test predictions for multi-wavelength emission that is associated with but not directly caused by FRBs. For example, we find evidence that the volumetric rate of magnetar giant flares that emit a factor of $\eta = 10^7$ larger energies in $\gamma$-rays than FRBs do in the radio band is over an order of magnitude higher than the FRB volumetric rate (Section~\ref{subsubsec:sgrs}). 
    
    \item The apparent rarity of multiwavelength FRB counterparts, and correspondingly likely low values of $\eta$, implies that future multiwavelength surveys are likely to only detect counterparts to the brightest FRBs. Given the steepness of the FRB fluence distribution at the  bright end \citep{2019MNRAS.483.1342J}, future blind surveys searching for FRB counterparts should prioritize sensitivity, and the ability to probe appropriately short timescales, over field of view and survey duration. 
    
\end{enumerate}

Although our results are robust to uncertainties in the FRB fluence distribution, the future application of our technique to better constrain FRB models will require a careful analysis of unclassified transient events in existing survey data sets. In addition, we assume a homogeneous population of FRB sources, which may not be the case, and it is also possible that some (e.g., soft X-ray, or blue optical) FRB counterparts are absorbed or scattered in dense surrounding media. Multi-wavelength observations of nearby individual sources \citep[e.g.,][]{2020ApJ...890L..32C,2020arXiv200312748P,2020arXiv200403676T} are a complementary means to address the nature of the FRB engine and emission mechanism.


\acknowledgments
We thank Sterl Phinney and Casey Law for useful discussions. 
This material is based upon work supported by the National Science Foundation under Grant No. AST-1836018. 
WL is supported by the David and 
Ellen Lee Fellowship at Caltech.



\bibliography{luminosity_bib}{}
\bibliographystyle{aasjournal}

\end{document}